\documentclass[12pt]{article}

\usepackage{epsfig}
\usepackage{amssymb}
\usepackage{amsfonts}

\usepackage{color}

\def\be{\begin{equation}}
\def\ee{\end{equation}}
\def\ba{\begin{array}{c}}
\def\ea{\end{array}}

\newcommand{\bea}{\begin{eqnarray}}
\newcommand{\eea}{\end{eqnarray}}

\newcommand{\bbr}{\br\!\br}
\newcommand{\kkt}{\kt\!\kt}

\newcommand{\kt}{\rangle}
\newcommand{\br}{\langle}

\begin{document}

\titlepage


 \begin{center}{\Large \bf

Non-Hermitian-Hamiltonian-induced unitarity and
optional physical inner products in Hilbert space

  }\end{center}


 \begin{center}

\vspace{8mm}

  {\bf Miloslav Znojil} $^{1,2}$

\end{center}

\vspace{8mm}

  $^{1}$
 {The Czech Academy of Sciences,
 Nuclear Physics Institute,
 Hlavn\'{\i} 130,
250 68 \v{R}e\v{z}, Czech Republic, {e-mail: znojil@ujf.cas.cz}}


 $^{2}$
 {Department of Physics, Faculty of
Science, University of Hradec Kr\'{a}lov\'{e}, Rokitansk\'{e}ho 62,
50003 Hradec Kr\'{a}lov\'{e},
 Czech Republic}

\newpage

\section*{Abstract}

For unitary quantum systems an innovative
construction of different evolution patterns
sharing the same energy-representing quasi-Hermitian
Hamiltonian $H\neq H^\dagger$ is proposed.
The differences originate from
the use of non-equivalent
physical inner product metrics
$\Theta=\Theta_\rho(H)$ with $\rho=0,1,2,\ldots$.
The main message is that
a weakening of the
isotropy of the Hilbert-space geometry
can be compensated by
a prolongation of
the interval of unitarity of the evolution.
Some of the related technical details
are illustrated via a
two-level and a four-level
exactly solvable toy models.

\subsection*{Keywords}.

Non-Hermitian quantum mechanics of closed systems;

Ambiguity of the physical inner product in Hilbert space;

Simplified subfamily of inner-product metrics;

Solvable two-state and four-state toy models;

\newpage

\section{Introduction}

Many years ago Freeman Dyson \cite{Dyson}
made a counterintuitive observation
that
even for certain manifestly unitary quantum systems
a mathematically optimal
representation of the Hamiltonian
may be non-Hermitian, $H \neq H^\dagger$.
At present, such an idea
is already incorporated in
several upgraded
formulations of quantum theory \cite{Geyer,Carl,SIGMA,ali,book}.

Naturally, Dyson
was well aware of the fact that
the concept of Hermiticity is
Hilbert-space-dependent
(or, more precisely,
inner-product-dependent)
so that he made it clear that
the non-Hermiticity property
of his Hamiltonians is
observed only
when one employs the most elementary Dirac's bra-ket  inner
product $\br \psi_a|\psi_b\kt$.
Characterizing the maximally
user-friendly
Hilbert space ${\cal H}_{math}$
which was, in the Dyson's models, unphysical.

The Dyson's followers
(cf., e.g., Scholtz et al \cite{Geyer})
emphasized that
another, correct and standard physical
Hilbert space ${\cal H}_{phys}$
can comparatively easily find its
representation in
the user-friendlier space
${\cal H}_{math}$,
provided only that one properly amends the inner product,
 \be
 \br \psi_a| \psi_b \kt \
 \to \
 \br \psi_a|\Theta\,| \psi_b \kt\,.
 \label{phyr}
 \ee
A key advantage of such a representation
(i.e., of
an innovated formulation of quantum mechanics (QM)
called quasi-Hermitian (QH))
lies in the ease with which
the obligatory self-adjointness of $H$ in ${\cal H}_{phys}$
appears equivalent to the $\Theta-$quasi-Hermiticity
property
 \be
 H^\dagger\,\Theta=\Theta\,H\,
 \label{thetaquia}
 \ee
of the same operator in
its preferred
representation in ${\cal H}_{math}$.

In our present letter we felt inspired
by the fact that once we admit an arbitrary
nontrivial choice of metric $\Theta \neq I$ in (\ref{thetaquia}),
we obtain a new tool of control of the dynamics of the
underlying physical
system.
This opens, in particular,
the possibility of a
unitary-evolution-mediated
access to the various new forms of singularities
(including, first of all the Kato's exceptional points \cite{Kato})
and to the related phase transitions and
to the Thom's
catastrophes \cite{Thom}
of a genuine quantum origin
\cite{catast}.

We will
outline the basic features of the QH QM formalism in
introductory section
\ref{quius},
pointing out that in most of its applications
a key to success lies in the Dieudonn\'{e}'s
quasi-Hermiticity relation
(\ref{thetaquia})
and in the related assignment of
a suitable and tractable ``physical'' inner-product metric $\Theta$
to a preselected and, in general, non-Hermitian
Hamiltonian $H$.

As long as the later assignment is well known
to be ambiguous \cite{Geyer,SIGMAdva},
people usually search for a sufficiently persuasive
method of suppression of the ambiguity
(cf., e.g., \cite{Geyer,lotor}).
In our present letter we decided to advocate
an alternative model-building philosophy
by which one starts from a larger multiplet of alternative
eligible inner-product metrics $\Theta$.

For the sake of definiteness,
we will number these metrics
by an auxiliary
integer subscript $\rho$:
The most
impatient readers may also
jump now to their
present, Hamiltonian-dependent
definition by Eq.~(\ref{psoucin}) below.
Needless to add, our choice of such a specific definition
was just methodical and, up to a large extent, arbitrary.
What is only important is that it leads
to an exceptionally user-friendly
construction
of the candidates for the metric, with
its ultimate unique selection to be
based on some additional {\it ad hoc\,} criteria.

For an explicit illustration of our proposal,
two versions of a toy model
will be then defined and studied in
sections \ref{trojacka}, \ref{interim} and \ref{sectionone},
with a few concluding comments
added in section \ref{huius}.

\section{Dyson maps\label{quius}}



In the stationary quasi-Hermitian quantum mechanics of review \cite{Geyer}
the description of
the bound states of a closed system
works with a pair of Hilbert
spaces, viz., ${\cal H}_{math}$ (which is unphysical) and ${\cal H}_{phys}$
(which is just represented in  ${\cal H}_{math}$ via amended inner product).
Often (cf. review \cite{Geyer}),
the available information about the underlying dynamics
(i.e., about the set of observables under consideration)
is assumed incomplete.
Still,
such an information
may prove ``sufficient for some
calculational purposes'' (cf. p. 74 in \cite{Geyer}),
especially when one is able to
factorize the physical inner-product metric,
 \be
 \Theta=\Omega^\dagger \Omega \neq I\,.
 \label{equus}
 \ee
The less usual non-unitary factor $\Omega$ called Dyson map
or, in \cite{Ju2022}, ``generalized vielbein'',
``if it exists'' \cite{Geyer},
has only to be invertible
since its basic role is to allow us to
reinterpret the
Hamiltonian
 $
 H\neq H^\dagger
 $
as a
Hermitizable operator,
i.e., as a preconditioned isospectral version of its conventional
self-adjoint
avatar
 \be
 \mathfrak{h}=\Omega\,H\,\Omega^{-1}
 =
 \mathfrak{h}^\dagger\,.
 \label{esa}
 \ee
The Hermiticity of $\mathfrak{h}$
(responsible for the unitarity of evolution \cite{Stone})
can be then re-read as equivalent to the quasi-Hermiticity
rule (\ref{thetaquia}) as imposed upon $H$.

In the literature
the terminology did not stabilize
yet.
Different authors speak not only
about the
quasi-Hermiticity of $H$ \cite{Geyer,Dieudonne}
but also about its
spontaneously unbroken
parity times time symmetry \cite{Bender1998,Ashida2020}
or about its pseudo-Hermiticity
\cite{ali,Mostafazadeh2002,Mostafazadeh2006} or
parity-charge-time symmetry \cite{Carl}
or crypto-Hermiticity \cite{ps}.
In all of these cases
the use of the factorization formula (\ref{equus})
and of the Dyson mapping of Eq.~(\ref{esa})
is well
motivated by the possibility of reference
to the principle of correspondence \cite{Messiah}.

In applications, the reconstruction
of the Dyson map {\it alias\,} vielbein $\Omega$
and/or of the self-adjoint partner Hamiltonian
$\mathfrak{h}$
of Eq.~(\ref{esa})
are not always needed.
Their genuine necessity
only emerges in the non-stationary
quasi-Hermitian models in which
even the metric itself is allowed to be time-dependent
(see, e.g., the outlines of the corresponding consistent
quantum theory in \cite{timedep,Fring}).
Nevertheless,
even in the
non-stationary scenario (a.k.a. non-Hermitian interaction picture,
NIP, cf. its reviews \cite{SIGMA,NIP,Bishop,Ju2019,Ju2024})
it makes sense to keep working with the
``instantaneous energy''
observable
$H=H(t)$ called Hamiltonian.

\subsection{Eligible physical inner product metrics}

An advantage of the latter decision
is that the reconstruction of
the
Dyson map can be then based on relation (\ref{esa}).
One can pick up a special, diagonal
form of the Hermitian matrix
$\mathfrak{h}=\mathfrak{h}_{0}^{}$
with the real (and, say, positive)
eigenvalue elements  $E_1, E_2,\ldots, E_N$.
This yields
the
zero-subscripted special case of
Eq.~(\ref{esa}),
 \be
 H^\dagger\,{\Omega}^\dagger_{0}={\Omega}^\dagger_{0}\,
 \left[ \begin {array}{cccc}E_1&0&\ldots&0
 \\\noalign{\medskip}0&E_2&\ddots&\vdots
 \\\noalign{\medskip}\vdots&\ddots&\ddots&0
 \\\noalign{\medskip}0&\ldots&0&E_N
 \end {array} \right]\,.
 \label{uieNu}
 \ee
This relation can be re-read as an $N$-plet of
auxiliary
Schr\"{o}dinger equations,
 \be
 H^\dagger\,|\psi_n\kkt = E_n\,|\psi_n\kkt\,,
 \ \ \ \ n=1,2,\ldots\,,N\,
 \label{qdirSE}
 \ee
which define and $N-$plet of
arbitrarily normalized ``ketket'' eigenvectors.
The
zero-subscripted
solution of Eq.~(\ref{uieNu}) can be then
reinterpreted as an $N$ by $N$ matrix composed of the
column-vector solutions $|\psi_j\kkt$
of Eq.~(\ref{qdirSE}),
 \be
 {\Omega}^\dagger_{0}=
 \left[ \begin {array}{cccc}
 |\psi_1\kkt\,,&|\psi_2\kkt\,,&\ldots&|\psi_N\kkt\,
 \end {array} \right]\,.
 \label{cuieNu}
 \ee
In the next step, in
a way pointed out in our older comment \cite{SIGMAdva} every
column-vector solution of Eq.~(\ref{qdirSE})
can be pre-multiplied by an arbitrary $n-$dependent constant.
This yields another acceptable Dyson map,
 \be
 {\Omega}^\dagger_{0}\ \to\  {\Omega}^\dagger(\vec{\kappa})=
 \left[ \begin {array}{cccc}
 |\psi_1\kkt\,\kappa_1,&|\psi_2\kkt\,\kappa_2,&\ldots&|\psi_N\kkt\,\kappa_N
 \end {array} \right]\,
 \label{kacuieNu}
 \ee
as well as  another acceptable metric,
 \be
 \Theta_{0}=\Omega^\dagger_{0}\Omega_{0}
 \ \to\
 \Theta(\vec{\kappa})=\Omega^\dagger(\vec{\kappa}) \Omega(\vec{\kappa})
 =\sum_{n=1}^N\,|\psi_n\kkt\,\kappa_n^2\,\bbr \psi_n|
 \,.
 \label{dequus}
 \ee
This formula defines {\em all\,} of the eligible inner product metrics
which are compatible with the preselected Hamiltonian.

A weakness of the latter result is that
formula (\ref{dequus})
is difficult to implement in practice,
mainly because the vectors $|\psi_j\kkt$ are not mutually orthogonal.
This is a difficulty which played the role of
a motivation of our present study.

\subsection{Hamiltonian-generated metrics}

The construction
of {\em any\,} acceptable metric
represents one of the key technical challenges during
a consistent quasi-Hermitian model-building process \cite{3a}.
In this sense
we found a sufficiently strong
encouragement in paper \cite{Ruzicka}
in which it has been
revealed that whenever a prescribed
Hamiltonian $H$ is $\Theta_{0}-$quasi-Hermitian,
it is also $\Theta_\rho-$quasi-Hermitian, with
 \be
 {\Theta}_\rho=\Theta_{0}\,H^\rho\,
 \label{psoucin}
 \ee
(notice that symbol $H^\rho$
denotes here the $\rho-$th power of matrix $H$).

In what follows
we are going to study the
consequences of the availability of the
latter special metrics with,
say, integer exponents $\rho =  1, 2, \ldots$.
The related alternative quantum models
numbered by $\rho$
share the same Hamiltonian but
still
acquire
different physical properties and probabilistic
interpretations
determined by the respective Hilbert-space
metrics
 \be
 \Theta_\rho (\vec{\kappa})=
 \Theta_0 (\vec{\kappa})\,H^\rho
 =\sum_{n=1}^N\,|\psi_n\kkt\,\kappa_n^2\,
 \left(E_n\right)^\rho\,\bbr \psi_n|
 \,.
 \label{ddequus}
 \ee
These matrices
have just the form of a reparametrized Eq.~(\ref{dequus}), and
they degenerate back to
$\Theta_{0}$ at $\rho=0$.
Nevertheless, as long as the eigenvectors $|\psi_n\kkt$
of $H^\dagger$ (cf. (\ref{qdirSE}))
are not mutually orthogonal,
the changes of the physical Hilbert-space geometry
as specified by Eq.~(\ref{ddequus})
remain strongly model-dependent.

\subsection{A brief note on evolution equations}

The study of unitary quantum systems
using several alternative inner-product spaces
(Hilbert spaces)
is a new model-building tendency in which
the conventional work with
self-adjoint
observables is extended to
include also the observables in their
non-Hermitian or, more precisely, quasi-Hermitian
representation (see, once more, the introduction
in the field in \cite{Geyer}).

In the corresponding quasi-Hermitian formulations
of the unitary quantum mechanics
one has to distinguish between the models in which
the underlying Dyson map is time-independent
(one can then speak about quantum mechanics in quasi-Hermitian
Schr\"{o}dinger picture \cite{ali})
or manifestly time-dependent.
The difference between
the two scenarios
involves, first of all,
the description of the evolution of the
states and of the wave functions.
In the former case, indeed,
such a description can be based on the mere non-Hermitian
generalization
 \be
 {\rm i}\partial_t \,|\psi(t)\kt = H\,|\psi(t)\kt\,,
 \ \ \ \ H \neq H^\dagger
 \label{msrek}
 \ee
of the conventional Schr\"{o}dinger
equation in which the (usually, just stationary)
generator of the evolution in time is
still the same operator as the operator which represents
the observable energy of the system.
In such a case, naturally, a key technical problem lies, indeed,
in the reconstruction (\ref{phyr}) of one of the admissible
physical inner products.

An entirely different situation is encountered in the deeply
non-stationary
models characterized by the use of the
manifestly time-dependent inner-product
metrics $\Theta=\Theta(t)$.
In these cases
(cf., e.g., \cite{timedep})
it is necessary to
solve a modified and manifestly $\Omega(t)-$dependent
Schr\"{o}dinger evolution
equation
 \be
 {\rm i}\partial_t \,|\psi(t)\kt = G(t)\,|\psi(t)\kt\,,
 \ \ \ \ G(t)=H(t)-\Sigma(t)\,,\ \ \ \
 \Sigma(t)={\rm i}\Omega^{-1}(t)\partial_t\Omega(t)
 \label{nipmsrek}
 \ee
in which the spectra of both of the
so called Schr\"{o}dinger evolution generator $G(t)$
and of the
so called quantum Coriolis force $\Sigma(t)$
are, in general, neither real
nor even symmetric with respect
to the complex conjugation \cite{ps}.

In the resulting non-Hermitian interaction picture of
quantum mechanics \cite{NIP,Ju2019}
one encounters an enormous increase of the technical
difficulties: {\it pars pro toto\,} let us only mention
that in such a non-stationary setting the abstract
structure of the quantum theory of unitary systems
becomes fairly close to the abstract structure
of the Einstein's classical theory of gravity.

In the latter respect a recommended reading is paper
\cite{Ju2022}
in which an emphasis has been put upon
the
geometric interpretation of the
non-stationary quasi-Hermitian quantum mechanics,
including even the concept of the
curvature of the space
and/or the possibility of
the evolution along
the generalized non-time dimensions.

Although the latter remark already concerns the questions
which lie beyond the scope of our present letter,
they may still
offer a
guide towards a next-step extension of our present message
in which we recommend an enhancement of the applicability
of both of the stationary and non-stationary non-Hermitian formalisms
using the explicit constructions based on formula (\ref{psoucin}).

\section{Illustration: two-level toy-model \label{trojacka}}

Among all of the phenomenology-related challenges in QH QM
a central position is occupied by the
emergence of the
accessibility of an
exceptional-point singularity (EP, \cite{Kato})
and of the
ambiguity of the Hilbert-space geometry
in its vicinity
(cf. \cite{corridors}).
Both of these consequences of the
enhancement of the flexibility
of the Hamiltonians
can be mutually interrelated
in a way which is, in general,
difficult to
quantify.

A successful quantification of
at least some of the EP-related phenomena
may
profit from the exact solvability
of models.
For illustration purposes let us consider, therefore,
the quasi-Hermitian two-state
Hamiltonian matrix
 \be
 H=H(t)=\left[ \begin {array}{cc} 1&\sqrt {1-{t}^{2}}
 \\\noalign{\medskip}-\sqrt {1-{t}^{2}}&3\end {array} \right]
 \,
 \label{ta2}
 \ee
in which, for illustration purposes
(and with an optional manifest reference
to the evolution
equations (\ref{msrek}) and (\ref{nipmsrek})), the parameter $t$
may though need not be interpreted as time.

It is straightforward to verify that the
spectrum of energies
is, at any real parameter, real, $E_\pm =E_\pm(t) =2 \pm t$.
It is also obvious that  whenever $t^2>1$, the latter
observation is trivial because
in such cases
our Hamiltonian matrix is Hermitian.

\subsection{Minimally anisotropic physical inner product metric\label{3.1}}

In a way explained
in paper \cite{passage}
the survival of the reality of the spectrum
in the interval of non-Hermiticity (i.e., for $t^2 \leq 1$)
implies that under certain additional conditions
(see their list in \cite{Geyer})
our toy model (\ref{ta2}) could be
made Hermitian in the two-dimensional
Hilbert space ${\cal H}_{math}=\mathbb{R}^2$ endowed, say, with
a special physical inner-product metric
 \be
 \Theta=\Theta_{0}(t)=
\left[ \begin {array}{cc} 1&-\sqrt {1-{t}^{2}}
\\\noalign{\medskip}-\sqrt {1-{t}^{2}}&1\end {array} \right]\,.
\label{thema}
 \ee
One only has to verify the validity of the above-mentioned
mathematical
conditions. Among them, one of the most important ones
is that
the eigenvalues
$\theta_\pm=1 \pm \sqrt {1-{t}^{2}}$
of the candidate (\ref{thema})
for the nontrivial physical inner-product metric
must be real. This is true
iff $t^2\leq 1$.

Another constraint is that
the real and symmetric
matrix (\ref{thema}) must
remain positive definite.
This is true when $\theta_-=1 - \sqrt {1-{t}^{2}}>0$,
i.e., in the two disjoint intervals of
$t \in [-1,0)$ and of $t \in (0,1]$.
Thus,
we may conclude that
matrix (\ref{thema})
defines an acceptable physical Hilbert-space metric
in both of the latter subintervals.
Moreover, after the amendment (cf. Eq.~(\ref{phyr}) above),
the new and correct
physical inner product
can be classified as
minimally anisotropic
so that in the light of paper \cite{lotor}
the geometry of the physical Hilbert space is unique.

\subsection{Exceptional-point singularity}

A serendipitious phenomenological
property of our highly instructive
but still extremely elementary $N=2$ model (\ref{ta2}) + (\ref{thema})
is that the point $t=0$
represents an
impenetrable physical discontinuity.
In the limit $t \to 0$, indeed, the
conventional diagonalization
 \be
 H^{(2)}(t) \ \to \ \mathfrak{h}^{(2)}(t)=\left[ \begin {array}{cc}2+t&0
 \\\noalign{\medskip}0&2-t\end {array} \right]\,,\ \ \ t \neq 0
 \label{Diana}
 \ee
cannot be performed anymore.
Our Hamiltonian (\ref{ta2}) can only be
assigned a non-diagonal canonical
Jordan-block representation,
 $$
 H^{(2)}(t) \ \to \ 
   \left[ \begin {array}{cc} 2&1\\\noalign{\medskip}0&2\end {array}
 \right]\,,\ \ \ t = 0 \,.
 $$
The value
of $t=0$ appears to be the Kato's exceptional point
of the Hamiltonian.
Strictly speaking,
operator $H^{(2)}(t)$ ceases to be
tractable as a quantum
Hamiltonian at $t=0$.

Once we decide to treat our free parameter $t$ as time,
we might try to interpret its passage from its small negative values to
its small positive values
as a quantum phase transition.
As we already mentioned above, this makes a detailed study of
the vicinity of the EP instant extremely attractive.
A
consequent analysis of the situation
reveals
that
the ``negative-time'' and the ``positive-time''
Hamiltonians may be entirely different~\cite{passage}.
Consistently, we can only speak
about a
unitary evolution
before or after
the exceptional-point time $t=t^{(EP)}=0$.

In what follows
we will only consider the latter scenario,
keeping in mind its slightly enhanced
potential methodical relevance
in quantum cosmology \cite{BigBang,BigBangb}.
In this setting, the extreme simplicity of model (\ref{ta2})
represents a significant methodical advantage because
its study may still remain non-numerical.
We intend
to use this advantage and
to show that the exact solvability
enables us to discuss the existence and properties of different
versions of the unitary evolution
near $t=0$ via closed formulae.

In particular, using the different choices of the exponent $\rho$
in the metric of Eq.~(\ref{psoucin}) we will be able to
describe
multiple alternative forms of the
unfolding of the EP singularity
representing
the non-equivalent phenomenological evolution scenarios
associated with the same
Hamiltonian operator.


\section{Exceptional-point unfoldings\label{interim}}

 \noindent
In the conventional Hermitian quantum mechanics
of textbooks
people {\em always\,} choose just the trivial
physical inner-product metric $\Theta_{textbooks}=I$.
In such a context the above claim of
existence of multiple
non-equivalent phenomenological evolution scenarios
associated with the same
Hamiltonian operator would sound like a paradox.

In the unconventional, quasi-Hermitian theory framework
it was much more easy to persuade its users
that
the variability and ambiguity of the
physical and Hamiltonian-dependent
inner-product metric $\Theta=\Theta(H)$
(restricted {\em solely\,} by relation (\ref{thetaquia}))
is ubiquitous
and, in fact, independent even of the Hermiticity status
of $H$ in ${\cal H}_{math}$ (cf. \cite{Geyer}).
The
traditional concept of Hermiticity
is relevant just in ${\cal H}_{phys}$.
In the working space ${\cal H}_{math}$
its role is inherited by the quasi-Hermiticity.
In an amended paradigm,
a more important role becomes played by the
multiparametric nature
of metric (\ref{dequus}).

One of the most important merits of the latter formula
is its relation to the
conjugate Schr\"{o}dinger Eq.~(\ref{qdirSE}).
The limiting transition
to $t^{(EP)}=0$
makes the individual states $|\psi_n\kkt$
mutually linearly dependent
so that, as expected,
the
acceptability (viz., invertibility) of
the metric is destroyed as a byproduct \cite{Kato}.

At the small but non-vanishing $t \neq 0$
both of the latter two degeneracies
disappear.
At the positive
and growing
times $t>0$, in particular,
the
process of unfolding
of the EP singularity
becomes regular and standard, i.e.,
unitary in  ${\cal H}_{phys}$.
Our present attention will be
aimed at its
ambiguity aspects,
i.e., at its
alternative
realizations
based on the
easy-to-construct variability of metrics (\ref{psoucin})
numbered by the integer $\rho$.

\subsection{$\rho=1$}

At $N=2$ and at any not too large time
$t>0$ the influence of the variability of $\rho$
upon the geometry of the
physical Hilbert space ${\cal H}_{phys}$
is best characterized by the
difference between the two eigenvalues
of the metric.
In technical sense, great advantage is
provided by the smallness of the Hilbert-space dimension $N=2$
because this gives us an opportunity of evaluation of
the eigenvalues of metrics $\Theta_\rho(t)$
of Eq.~(\ref{psoucin})
at any $\rho$ in a compact analytic form.
The first nontrivial illustration is provided by the choice of
$\rho=1$ yielding the metric
 $$
 {\Theta}_1=
\left[ \begin {array}{cc} 2-{t}^{2}&-2\,\sqrt {1-{t}^{2}}
\\\noalign{\medskip}-2\,\sqrt {1-{t}^{2}}&2+{t}^{2}\end {array}
 \right]
 $$
with eigenvalues
 $$
 \theta_+=4-t^2\,,\ \ \
 \theta_-=t^2\,.
 $$
We see that in contrast to the trivial $\rho=0$
model which only existed at $|t|\leq 1$ and which
was discussed, rather thoroughly, in paragraph \ref{3.1} above,
the alternative $\rho=1$ eigenvalues remain real at all times $t>0$.
Moreover, they remain both positive,
i.e., compatible with the unitarity of the model in ${\cal H}_{phys}$
in a perceivably longer interval of time $t \in (0,2)$.

Even at the ``very large'' times
$t>2$ the operator $\Theta_1=\Theta_1(t)$ which
loses its positive definiteness still
retains its invertibility.
Such a matrix remains tractable as an indefinite pseudo-metric
in Krein space \cite{ali}.
In the literature, people also often speak about
${\cal PT}-$symmetric models in such a case
(see more details, e.g., in \cite{Christodoulides,Carlbook}).

\subsection{$\rho=2$}

Once we move to $\rho=2$
we get the metric
 $$
 {\Theta}_2=
 \left[ \begin {array}{cc} 4-3\,{t}^{2}&-4\,\sqrt {1-{t}^{2}}-{t}^{2}\,\sqrt {1
-{t}^{2}}\\\noalign{\medskip}-4\,\sqrt {1-{t}^{2}}-{t}^{2}\,\sqrt {1-{t}
^{2}}&4+5\,{t}^{2}\end {array} \right]
 $$
with eigenvalues
 \be
 \theta_\pm=4+{t}^{2} \pm \sqrt {16-8\,{t}^{2}+9\,{t}^{4}-{t}^{6}}\,.
 \label{tydve}
 \ee
The closed algebraic form of this result
enables us to conclude that the
system remains unitary in a still larger interval of time
$t \in (0,t_2)$ where $t_2\approx  2.875129794$.
The
value of the upper boundary
beyond which the metric eigenvalues complexify
can be
prescribed by the  {\em exact\,}
Cardano formula,
$$
3\,t_2=1+\sqrt [3]{73+6\,\sqrt {87}}+13\,/{{\sqrt [3]{73+6\,
\sqrt {87}}}}\,.
$$
Moreover, there emerges an analogy between the shapes
of the time-dependence
of the
spectra of metrics with the same parity of $\rho$:
At $\rho=2$ this spectrum becomes also complex
beyond $t_2$.

\subsection{$\rho=3$}

An analogous conjecture can be made for any odd
$\rho$ for which the smaller eigenvalue becomes negative
at some critical time $t=t_\rho$  (cf. the case of $\rho=1$ above).
At $\rho=3$, the survival of the comparatively elementary
closed form of the metric
 $$
 {\Theta}_3=
  \left[ \begin {array}{cc} 8-6\,{t}^{2}-{t}^{4}&-8\,\sqrt {1-{t}^{2}}-
 6\,{t}^{2}\sqrt {1-{t}^{2}}
 \\\noalign{\medskip}-8\,\sqrt {1-{t}^{2}}-6
 \,{t}^{2}\sqrt {1-{t}^{2}}&8+18\,{t}^{2}+{t}^{4}\end {array} \right]
 $$
leads to the compact closed-form eigenvalues
 $$
 \theta_\pm =8+6\,{t}^{2} \pm
  \sqrt {64+32\,{t}^{2}+84\,{t}^{4}-12\,{t}^{6}+{t}^{8}}\,.
  $$
which remain both positive in the interval of $t \in (0,2)$.

\subsection{$\rho \geq 4$}

At $\rho=4$ we have
 $$
 {\Theta}_4=
   \left[ \begin {array}{cc} 16-8\,{t}^{2}-7\,{t}^{4}&-16\,\sqrt {1-{t}^
{2}}-24\,\sqrt {1-{t}^{2}}{t}^{2}-\sqrt {1-{t}^{2}}{t}^{4}
\\\noalign{\medskip}-16\,\sqrt {1-{t}^{2}}-24\,\sqrt {1-{t}^{2}}{t}^{2
}-\sqrt {1-{t}^{2}}{t}^{4}&16+56\,{t}^{2}+9\,{t}^{4}\end {array}
 \right]
 $$
and
 $$
 \theta_\pm =16+24\,{t}^{2}+{t}^{4} \pm \sqrt {256+512\,{t}^{2}+864\,{t}^{4}-48\,{t}^{6
}+17\,{t}^{8}-{t}^{10}}
 $$
The growth of the interval of unitarity with $t\in (0,t_4)$
is accelerating since
$t_4\approx 4.150651137$. We see that the square of
this quantity is a root of a polynomial of the fifth degree
so that its representation via a closed formula is not expected
to exist anymore.

\begin{figure}[h]                     
\begin{center}                         
\epsfig{file=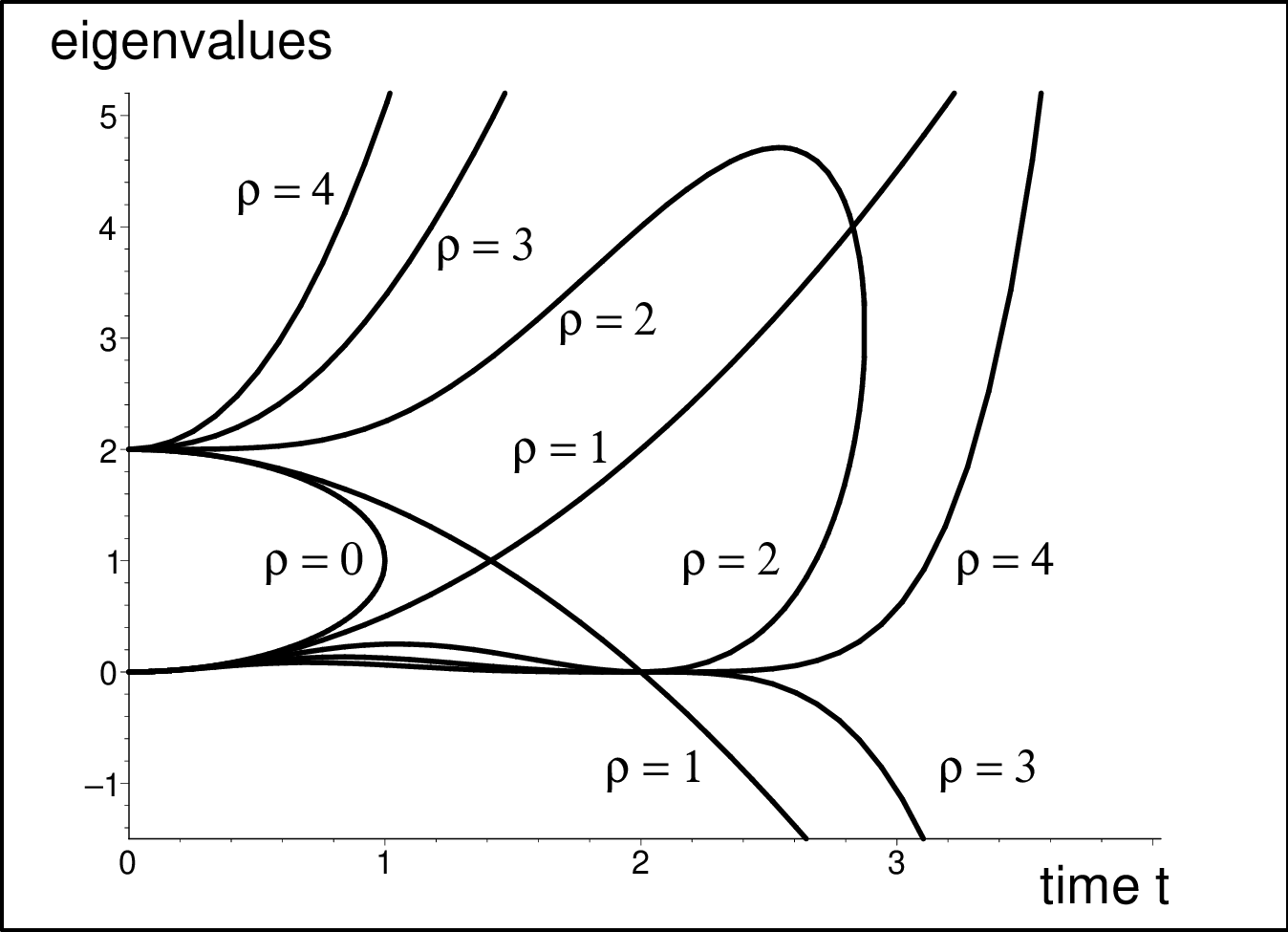,angle=270,width=0.35\textwidth}
\end{center}                         
\vspace{-2mm}\caption{Time-dependence of the (rescaled) spectra
of matrices $\Theta^{(2)}_\rho(t)$.}
 \label{zacal5x}
\end{figure}

A global numerical
summary of the time- and $\rho-$dependence
of the $N=2$ anisotropies of the inner products
is displayed in
Figure \ref{zacal5x}.
In the picture
we
rescaled
all metrics $\Theta_\rho \to C_\rho\,\Theta_\rho $
in such a way that
the spectra
coincide
in the EP limit.

The main message delivered by the picture
is that the
anisotropy of the
physical Hilbert-space inner product
grows with $\rho$ but that such a growth
remains very slow
in a small vicinity of $t=t^{(EP)}=0$.
Nevertheless,
as long as
the $\rho-$dependence of the anisotropy
becomes more pronounced at the larger times,
our attention becomes attracted to
a quantification of such a far-from-EP time-dependence.
Naturally, the choice of $N=2$
becomes too schematic for the purpose.

\section{Four-level model\label{sectionone}}


As long as the roots of the secular equations can be given by closed formulae
up to $N=4$, it makes sense to complement our preceding analysis
by its $N=4$ analogue using, say, another toy-model Hamiltonian
 \be
H^{(4)}(t)= \left[ \begin {array}{cccc} 1&\sqrt {3}\sqrt
{1-{t}^{2}}&0&0\\\noalign{\medskip}-\sqrt {3}\sqrt
{1-{t}^{2}}&3&2\,\sqrt {1-{t}^{2}} &0\\\noalign{\medskip}0&-2\,\sqrt
{1-{t}^{2}}&5&\sqrt {3}\sqrt {1-{t}^
{2}}\\\noalign{\medskip}0&0&-\sqrt {3}\sqrt {1-{t}^{2}}&7\end
{array}
 \right]\,.
 \label{BMW}
 \ee
As long as such a Hamiltonian represents just an inessential
modification of the
$N=4$ toy model
of paper \cite{passage},
we can follow this reference and
factorize the related secular polynomial
non-numerically. This yields the
explicit form of the energy spectrum
 $$
 \{E_n\} = \{ 4 - 3\,t, 4- t,  4+ t, 4 + 3\,t\}
 $$
which is real at any real time-parameter $t$
and positive for not too large $|t|<4/3$.

The choice of $N=4$ enables us to treat matrix
(\ref{BMW}) as bridging the gap between
our preceding oversimplified two-level model
and the large-$N$ matrices
of paper \cite{passage}
which could be perceived as
certain discrete equidistant-level analogues of the continuous
harmonic oscillator.

At $N=4$, in a close parallel to the two-level model
the real line of $t$ is again separated, by
its unphysical exceptional-point origin $t=t(^{(EP)}=0$,
into the negative-time
half-line of
the process of collapse,
and the positive-time
half-line of
the process of the EP unfolding.
Once we restrict again our attention to the half-line of $t>0$
we may still separate
the half-line of
$t\geq 1$
(where our Hamiltonian matrix becomes
manifestly Hermitian) from
the finite subinterval of our more immediate interest where
$t<1$, i.e., where our Hamiltonian matrix is
non-Hermitian but Hermitizable. i.e.,
Hermitian only in ${\cal H}_{phys}$.

\subsection{Inner product metric with $\rho= 0$\label{uysectionone}}

It is entirely routine to verify that our
Hamiltonian
is quasi-Hermitian
with respect to
the special metric
$\Theta_{0}^{(4)}(t)$ equal to matrix
 \be
 \left[ \begin {array}{cccc} 1&-\sqrt {3}\sqrt
{1-{t}^{2}}&\sqrt {3} \left( 1-{t}^{2} \right) &- \left( 1-{t}^{2}
\right) ^{3/2}
\\\noalign{\medskip}-\sqrt {3}\sqrt {1-{t}^{2}}&3-2\,{t}^{2}&-2\,
\sqrt {1-{t}^{2}}- \left( 1-{t}^{2} \right) ^{3/2}&\sqrt {3} \left(
1- {t}^{2} \right) \\\noalign{\medskip}\sqrt {3} \left( 1-{t}^{2}
 \right) &-2\,\sqrt {1-{t}^{2}}- \left( 1-{t}^{2} \right) ^{3/2}&3-2\,
{t}^{2}&-\sqrt {3}\sqrt {1-{t}^{2}}\\\noalign{\medskip}- \left(
1-{t}^ {2} \right) ^{3/2}&\sqrt {3} \left( 1-{t}^{2} \right) &-\sqrt
{3} \sqrt {1-{t}^{2}}&1\end {array} \right]\,.
\label{miiafa}
 \ee
This metric is, in the spirit of paper \cite{lotor}, minimally anisotropic.

This can be confirmed by the inspection
of its eigenvalues
obtainable as the roots of the related
secular polynomial of the fourth order.
We should only add that in this case the
exact solvability does not mean simplicity.
Indeed, the
display of the
analytic formulae for eigenvalues
would require many pages in print.
At the same time, the reliability of the formulae
is guaranteed since
they are all computer-generated by
the commercial
symbolic
manipulation software \cite{MAPLE}.

\begin{figure}[h]                     
\begin{center}                         
\epsfig{file=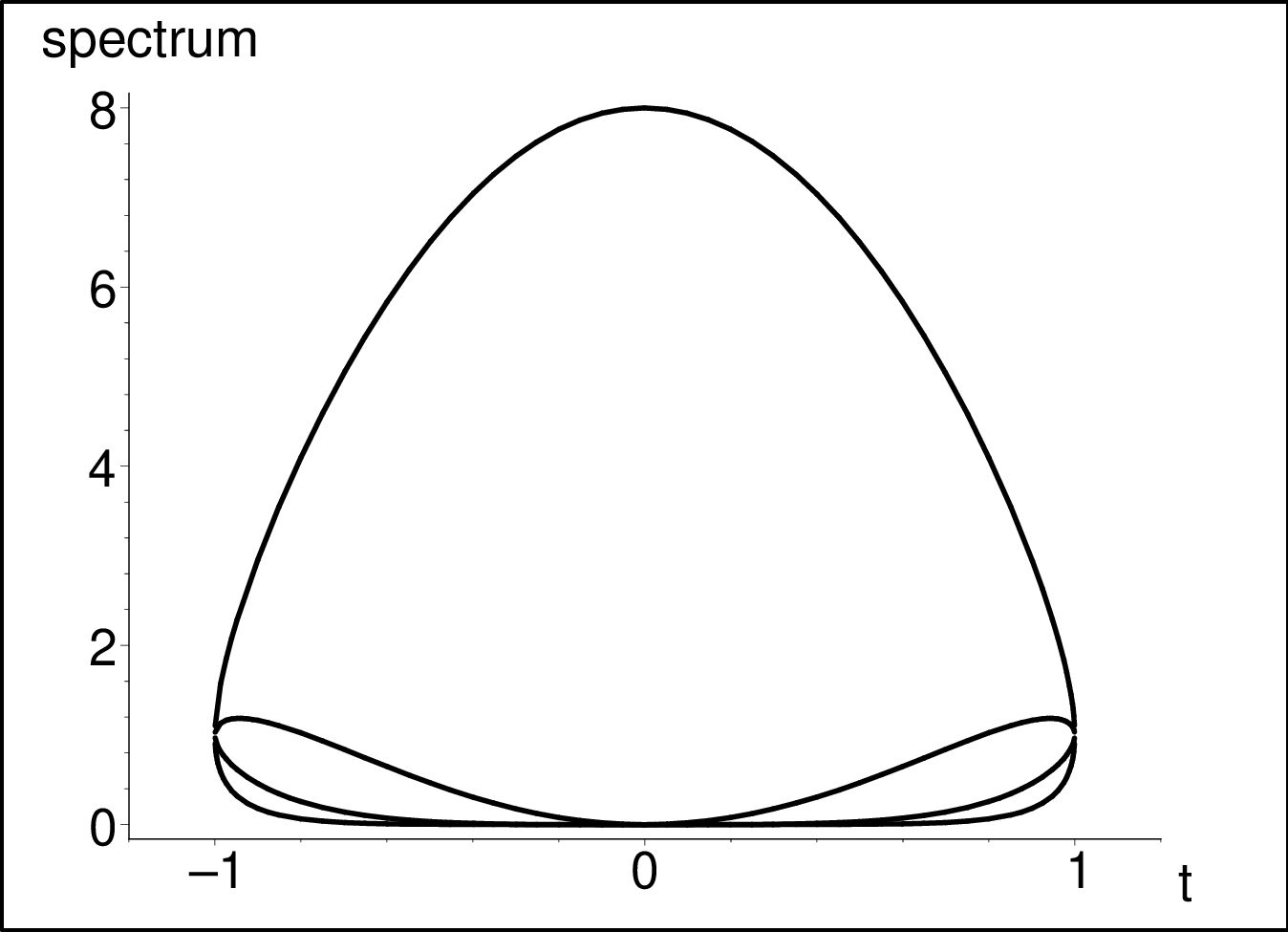,angle=270,width=0.35\textwidth}
\end{center}                         
\vspace{-2mm}\caption{Real eigenvalues of matrix~(\ref{miiafa}).
  }
 \label{zacal0}
\end{figure}
%

The inspection of  Figure \ref{zacal0} reveals that
in the vicinity of the origin (i.e., of the EP limit $t \to 0$)
an enhancement
of
the
numerical precision is important for the control of numerical errors.
Only then, indeed, a firm guarantee of the
positivity
of the smallest eigenvalue
(i.e., of the
unitarity of the evolution)
is obtained.
One can conclude that
all of the eigenvalues of metric $\Theta_{0}^{(4)}(t)$ are
positive and real if and only if
 $t \in [-1,0)$ or
  $t \in (0,1]$.
Their size
is strongly dominated by the single maximal eigenvalue
with its maximum (equal to 8) at $t=0$.

The strict isotropy of the geometry is only reached
at both of the limits of $t \to \pm 1$ at which the Hamiltonian
itself becomes diagonal. The same Hamiltonian
matrix remains
Hermitian whenever $|t| \geq 1$.
In the latter domain the
minimally anisotropic
metric degenerates to the identity operator.
In a complete analogy with the preceding $N=2$ scenario,
this feature makes again the
minimally anisotropic (i.e., $\rho=0$) geometry of the
physical $N=4$ Hilbert space
non-analytic at  $t = \pm 1$.

In the EP limit of $t \to 0$
the metric
loses its invertibility and
the operator degenerates to a matrix of rank one.
Figure \ref{zacal0} indicates that
its physical role and meaning are
lost
beyond
the points $t=\pm 1$.
The degeneracy of the spectrum
is
followed there by its complexification.
The quantum
system itself
ceases to be quasi-Hermitian
because its physical inner-product metric
ceases to be a Hermitian matrix.

Due to the left-right symmetry of the graph of Figure \ref{zacal0},
it makes sense again to consider
only the positive times $t \in (0,\infty)$ in what follows.

\subsection{Metrics with $\rho\neq 0$ \label{wuysectionone}}

The diagonalizations of
matrices with dimension $N\leq 4$
can be all classified as
non-numerical, in principle at least.
One only rarely encounters nontrivial
models with $N=4$ in which the use of
the
closed form of eigenvalues
would prove really useful. For practical purposes,
the $N=4$ formulae are too long and complicated in general.
Still, our $\rho-$numbered family of
models sharing the same Hamiltonian (\ref{BMW})
can be perceived as a
user-friendly exception in which the formulae
enable us to test the acceptability of the
candidates
$\Theta_{0}^{(4)}(t)$ for the metrics
using any necessary numerical level of precision.

On this ground we can
subdivide,
reliably, the real half-line of times $t \in (0,\infty)$
into subintervals  supporting
or  not supporting
the unitarity.
Methodically, we may feel guided by our
preceding $N=2$ results and by
the related spectra
as presented in Figure \ref{zacal5x}.
Several qualitative
features of the spectra
(represented now by the quadruplets of eigenvalues)
re-occur also at $N=4$.

%
%
\begin{figure}[h]                     
\begin{center}                         
\epsfig{file=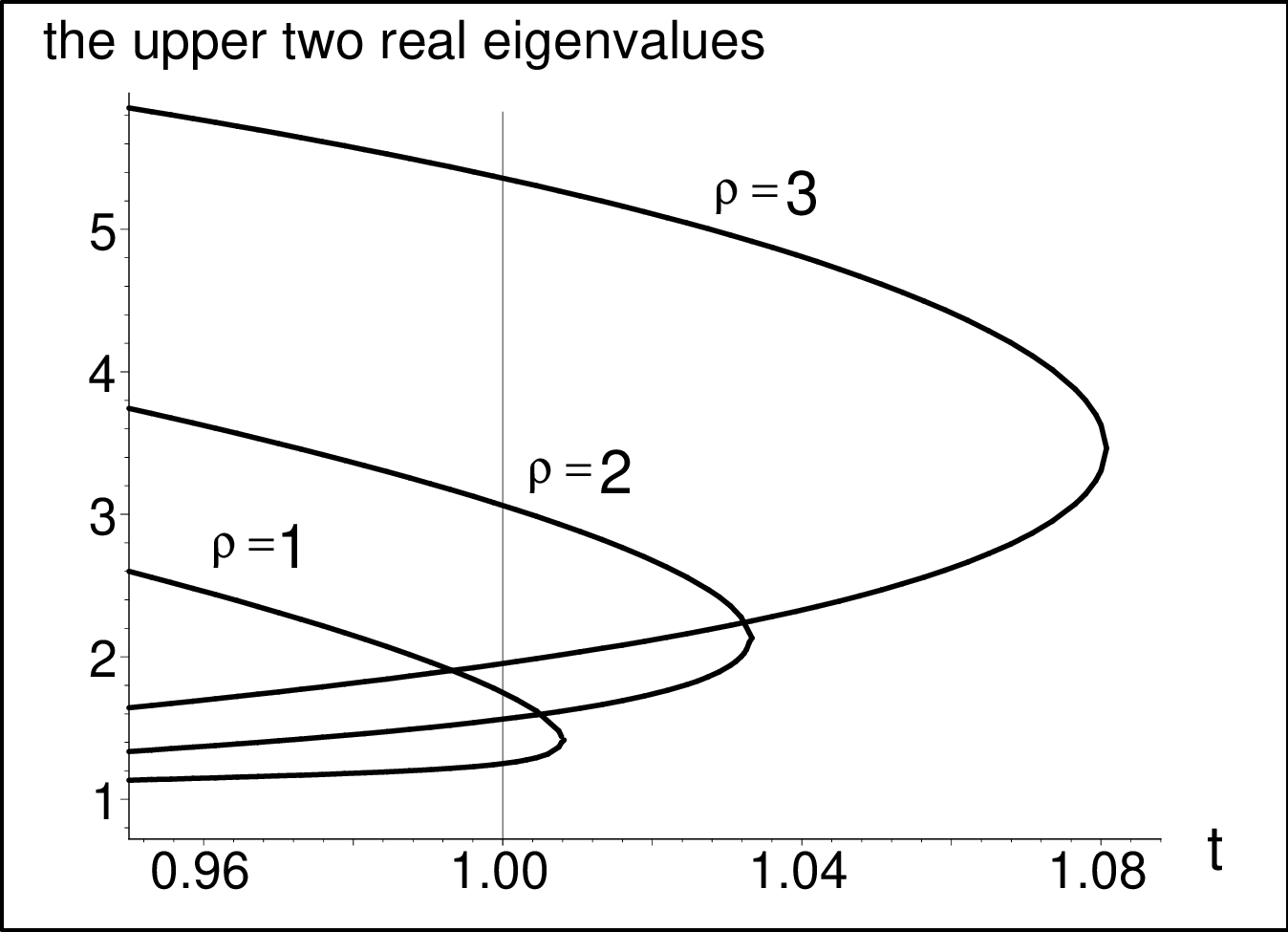,angle=270,width=0.35\textwidth}
\end{center}                         
\vspace{-2mm}\caption{The
two largest real eigenvalues of
matrices
$\Theta_\rho^{(4)}(t)$ with $\rho=1,2$ and $3$.}
 \label{sizacali}
\end{figure}

For the sake of the clarity of presentation we
introduce again a suitable $t-$independent
multiplicative factor $C_\rho$
and we
re-scale matrices
$\Theta_\rho^{(4)}(t)$
in such a way that
their spectra coincide in the EP limit $t \to 0$.
Then the calculations confirm that
at the not too large times $t \ll 1$
the
$\rho-$dependence of the spectrum remains weak,
i.e., the growth of the anisotropy
of the physical Hilbert space geometry
remains slow.

The situation becomes different at the larger times.
The detailed analysis is
guided by
our experience with $N=2$.
First of all, we detect again
a prolongation
of the $\rho-$dependent interval
$t \in (0,t_\rho)$  of quasi-Hermiticity
beyond its $\rho=0$ minimum with $t_0=1$.
Quantitatively, the phenomenon
is illustrated
in Figure \ref{sizacali}
where we omitted the two smaller
eigenvalues as irrelevant because both of them
remain safely real and, in the interval of interest,
also safely positive.

The picture clearly demonstrates that,
and how quickly,
the boundary $t=t_\rho$ of the consistency of the model
grows with the value of the exponent $\rho$
in the definition of metric~(\ref{psoucin}).
The spectrum of $\Theta_\rho (t)$ becomes complex
beyond the value of $t=t_\rho$
(marking the
$\rho-$dependent
end of the
quasi-Hermitian regime)
so that
the respective matrices
$\Theta_\rho^{(4)}(t)$ cease to be Hermitian at $t>t_\rho$.

One of the most remarkable consequences
of our turn of attention to the
models with nontrivial $\rho\neq 0$
(i.e., to the
corresponding alternative physical Hilbert-space geometries)
is the
prolongation of the
interval of times during which the system remains
(quasi-)Hermitian.
The weakening of the isotropy of the geometry
appears compensated by an extension
of the toy-model unitarity to the
larger values of time.


\section{Conclusions\label{huius}}

Having used, for methodical purposes,
two elementary solvable $N-$level quantum models
with $N=2$ and $N=4$,
we constructed
several
formally equally acceptable
modifications of the
underlying physical Hilbert-space
geometry.

In comparison with the conventional
unique choice of such a geometry
(characterized by a
maximally isotropic inner-product
metric $\Theta_0$)
we revealed that, unexpectedly,
a comparatively small increase of anisotropy
appeared compensated by
a prolongation of the interval $(0,t_\rho)$
of the time of the
existence and observability of the system.

One of the rather puzzling consequences of
our present 
exactly solvable toy-model constructions is that
in certain stages of unitary evolution
(i.e., at $t\in (1,t_\rho)$),
the Hamiltonian matrix $H$ representing the instantaneous
energy becomes Hermitian,
simultaneously, in the {\em two non-equivalent\,}
Hilbert spaces, viz., in the
``correct physical'' space ${\cal H}_{phys}$ and
in the ``auxiliary unphysical'' space
${\cal H}_{math}$.

By construction, the respective geometries
of these two spaces are
non-equivalent, but
only the former one
can be given the standard probabilistic interpretation
yielding consistent
unitary quantum systems
with
specific measurable characteristics.
Paradoxically, the
second space and geometry are, in spite of their simpler form,
spurious.

\section*{Acknowledgement}

The author was supported by
Faculty of Science of
University of Hradec Kr\'{a}lov\'{e}

\end{document}